\documentclass[twocolumn,showpacs,preprintnumbers,amsmath,amssymb]{revtex4}

\usepackage{graphicx}
\usepackage{dcolumn}
\usepackage{bm}
\usepackage{epsfig}
\usepackage{psfrag}
\usepackage{subfigure}

\begin{document}


\title{Layered Complex Networks}

\author{Maciej Kurant}
\author{Patrick Thiran}
 \affiliation{EPFL, Switzerland}

\begin{abstract}
Many complex networks are only a part of larger systems, where a number of coexisting topologies interact and depend on
each other. We introduce a layered model to facilitate the description and analysis of such systems. As an example of
its application we study the load distribution in three real-life transportation systems, where the lower layer is the
physical infrastructure and the upper layer represents the traffic flows. This layered view allows us to capture the
fundamental differences between the real load and commonly used load estimators, which explains why these estimators
fail to approximate the real load.
\end{abstract}


\maketitle

The topologies of the Internet at the IP layer~\cite{Faloutsos99}, of the World Wide Web~\cite{Albert99} or the
networks formed by Peer To Peer (P2P) applications~\cite{Adamic01} have recently drawn a lot of attention. These graphs
have been studied separately, as distinct objects. However, they are closely related: each WWW or P2P link virtually
connects two IP nodes. These two IP nodes are usually distant in the underlying IP topology and the virtual connection
is realized as a path found by IP routers. In other words, the graph formed by an application is mapped on the
underlying IP network. Although the topologies at both layers might share a number of statistical properties (such as a
heavy-tailed degree distribution), they are very different.\\
There exist layers also under the IP layer; even in a simplified view of the Internet we must distinguish at least one
- the physical layer. It consists of a mesh of optical fibers that are usually put in the ground along roads, rails, or
power-lines. This results in topologies very different from those observed at the IP layer. A mapping of the IP graph
onto the physical layer must satisfy a number of constraints (see e.g.,~\cite{Kurant05}).

Another important class of real-life systems is transportation networks. The graphs based on the physical
infrastructure of such networks were analyzed on the examples of a power grid \cite{Watts98,AlbertUSAPowerGrid},
railway network \cite{Indian03}, road networks \cite{Gastner04,Rosvall05}, or urban mass transportation systems
\cite{Latora01,Stations04,Gastner04b,Sienkiewicz05}. Although this approach often gives a valuable insight into the
studied topology, it ignores the real-life traffic pattern and hence captures only a part of the full picture.
Interestingly, the networks of traffic flows were studied separately, for instance the flows of people within a
city~\cite{Chowell03}, and commuting traffic flows between different cities~\cite{Montis05}. These studies, in turn,
neglect the underlying physical topology. A comprehensive view of the system requires to analyze both layers (physical
and traffic) together. Of course, a partial knowledge of the traffic pattern could be introduced into the physical
graph by assigning weights reflecting the amount of carried traffic to the physical edges. This describes well a
specific type of transportation network, where all traffic flows are one-hop long and where the two layers actually
coincide, such as airport networks~\cite{Barrat04,Guimera05}. However, in the presence of longer (than one hop) traffic
flows, the weighted physical graph is not sufficient. For instance, the failure of a single physical node/edge should
affect (delete or cause to reroute) all traffic flows using this edge/node, which requires the knowledge of the traffic
graph and of the actual routes of these flows in the physical graph.

\begin{figure*}[!t]
    \psfrag{GP}[][]{$G^\phi$}
    \psfrag{GL}[][]{$G^\lambda$}
    \psfrag{Gpb}[][]{{\large$G^\phi$}}
    \psfrag{Glb}[][]{{\large$G^\lambda$}}
    \psfrag{M}[][]{$\;M(\!E^\lambda\!)$}
    \psfrag{A}[][]{a)}
    \psfrag{B}[][]{b)}
    \psfrag{El1}[][]{\footnotesize{$e^\lambda_1$}}
    \psfrag{V1}[][]{\footnotesize{$v^\phi_1$}}
    \psfrag{V2}[][]{\footnotesize{$v^\phi_2$}}
    \psfrag{V3}[][]{\footnotesize{$v^\phi_3$}}
    \psfrag{M1}[][]{\footnotesize{$M(\!e^\lambda_1\!)$}}
\includegraphics[width=1\textwidth]{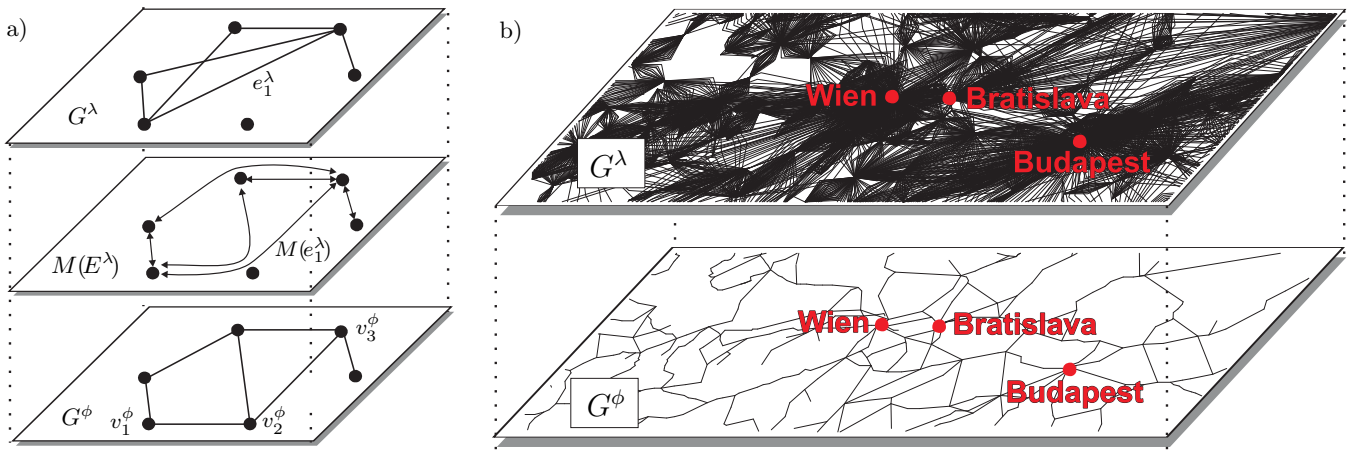}
\\[-0.25cm]
\caption{(a) Illustration of the two-layered model. The logical graph~$G^\lambda$ is mapped onto the physical graph~
$G^\phi$ by a mapping~$M(E^\lambda)$. In this example the logical edge $e^\lambda_1$ is mapped on $G^\phi$ as the path
$M(e^\lambda_1)=(v^\phi_1, v^\phi_2, v^\phi_3)$. Assuming that~$G^\lambda$ is unweighted, the loads of the three
indicated nodes are $l(v^\phi_1)\!=\!3$, $l(v^\phi_2)\!=\!2$ and $l(v^\phi_3)\!=\!4$. \; (b) A part of the logical and
physical graphs of the EU dataset. Here, the traffic intensities (weights) are indicated by multiedges in the logical
graph.\\[-0.7cm]} \label{fig:Example}
\end{figure*}

Coexisting and dependent graphs can also be observed in social networks~\cite{Wasserman94}, where the same set of nodes
may be connected in various ways, depending on the type of relationship chosen to be represented by edges. These graphs
are related to each other. It is common, for instance, to establish a new link in a business relationship graph (e.g.,
to find a job) by performing a search in our acquaintanceship network (i.e., by asking our friends who ask their
friends, etc)~\cite{Granovetter73}. This new direct business link translates into a path in the acquaintanceship
network.

The above examples call for the introduction of additional layers to the description of some complex systems. Therefore
we propose a general \emph{multi-layered model}. We explain it on the example of two layers; all the definitions
naturally extend to any number of layers. In the two-layered model, the lower-layer topology is called \emph{physical
graph} $G^\phi=(V^\phi, E^\phi)$, and the upper-layer topology is called \emph{logical graph} $G^\lambda=(V^\lambda,
E^\lambda)$. We assume that the sets of nodes at both layers are identical, i.e., $V^\phi\equiv V^\lambda$, but as a
general rule we keep the indexes $\phi$ and $\lambda$ to make the description unambiguous. Let $N$ be the number of
nodes, $N\!\!=\!\!|V^\phi|\!\!=\!\!|V^\lambda|$. The physical and logical graphs can be directed or undirected,
depending on the application. The nodes and edges can have \emph{weights} assigned to them and denoted by $w(\cdot)$,
with $w=1$ for unweighted graphs. Every logical edge $e^\lambda=(u^\lambda,v^\lambda)$ is mapped on the physical graph
as a path $M(e^\lambda)\subset G^\phi$ connecting the nodes~$u^\phi$ and~$v^\phi$, corresponding to $u^\lambda$ and
$v^\lambda$. (A path is defined by the sequence of nodes it traverses.) The set of paths corresponding to all logical
edges is called \emph{mapping} $M(E^\lambda)$ of the logical topology on the physical topology. Now, the \emph{load}
$l$ of a node
$v^\phi$ is the sum of the weights of all logical edges whose paths traverse~$v^\phi$:\\[-0.3cm]
\begin{equation}\label{eq:node_load}
l(v^\phi)=\sum_{e^\lambda\!:\;v^\phi\in M(e^\lambda)} w(e^\lambda)\\[-0.1cm]
\end{equation}
In a transportation network $l(v^\phi)$ is the total amount of traffic that flows
through the node $v^\phi$; if the logical graph is unweighted, $l(v^\phi)$ counts the number of logical edges that are
mapped on $v^\phi$.

Here, we apply this two-layered framework to study transportation networks. The undirected, unweighted physical graph
$G^\phi$ will henceforth capture the physical infrastructure of a transportation network, and the logical graph
$G^\lambda$ will reflect the undirected traffic flows. All data studied in this paper is extracted from timetables of
public transportation systems. First, we take a list of all of trains, metros and buses departing in the system within
one weekday (time-span of 24 hours). A timetable gives the exact route of each vehicle, which translates directly into
a logical edge $e^\lambda$ (connecting the first and the last station) and its mapping $M(e^\lambda)$. The number of
vehicles following the same path in both possible directions defines the flow intensity - the weight $w(e^\lambda)$ of
the logical link.
(In this context, the logical graph is equivalent to a traffic matrix in transportation
science~\cite{TransportationBook}.) We describe the algorithm to extract the two layers and the mapping from timetables
in~\cite{KurantRailwayAlgorithm}.

\begin{figure}[!b]
${}$\\[-0.5cm]
\psfrag{x1}[][]{\small{$k^\phi$}} \psfrag{y1}[][]{{\small $P(k^\phi)$}} \psfrag{x2}[][]{{\small$k^\lambda$}}
\psfrag{y2}[][]{{\small $P(k^\lambda)$}} \psfrag{x3}[][]{{\small$w^\phi$}} \psfrag{y3}[][]{{\small$P(w^\phi)$}}
\psfrag{x4}[][]{{\small$h$}} \psfrag{y4}[][]{{\small$P(h)$}} \psfrag{A}[][]{a)} \psfrag{B}[][]{b)} \psfrag{C}[][]{c)}
\psfrag{D}[][]{d)} \psfrag{real}[][]{{\footnotesize {}\;real}} \psfrag{betweenness}[][]{{\footnotesize all-to-all}}
\includegraphics[width=0.45\textwidth]{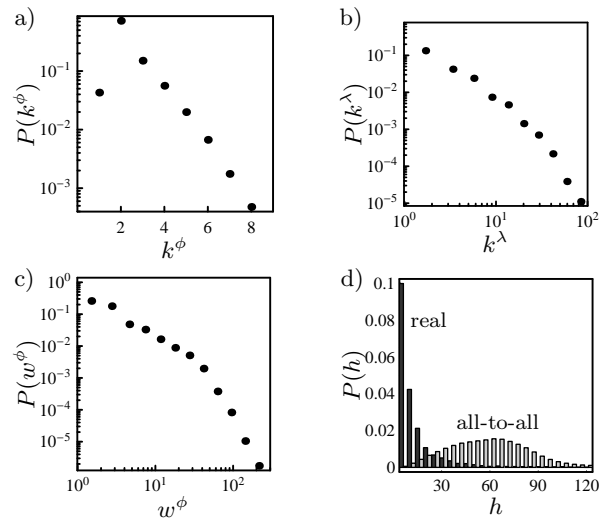}\\[-0.2cm]
\caption{EU network (WA and CH yield similar results). Node degree distribution in the physical graph~(a), and in the
logical graph~(b); edge weight distribution in the logical graph~(c); and the distribution of the lengths of traffic
flows~(d), counted in a number of hops $h$ in the physical graph.} \label{fig:Flows}
\end{figure}

\begin{table}[!htb]
\begin{tabular}{|l|c|c|c|c|c|c|c|}
  \hline
  Dataset & $N$ & $|E^\phi|$&$d^\phi$&$|E^\lambda|$&\# vehicles\\
  \hline
  WA - Warsaw & 1529 & 1827 & 90  &324& 26075\\
  CH - Switzerland& 1679 & 1750  & 142  & 539&7482\\
  EU - Europe &6276& 7273 & 181 & 6623& 54073\\
  \hline
\end{tabular}
\caption{The studied datasets. $N$ is the number of nodes, $|E^\phi|$ (respectively, $|E^\lambda|$) is the number of
edges in the physical (respectively, logical) graph, and $d^\phi$ denotes the diameter of the physical graph.
The total number of vehicles taken into account for every dataset is given by ``\# vehicles''. Note
that $|E^\lambda|\ll$~\# vehicles, because many vehicles follow the same route.\\[-0.5cm]}
\label{tab:Data}
\end{table}
\begin{figure*}[!htb]
 \includegraphics[width=1\linewidth]{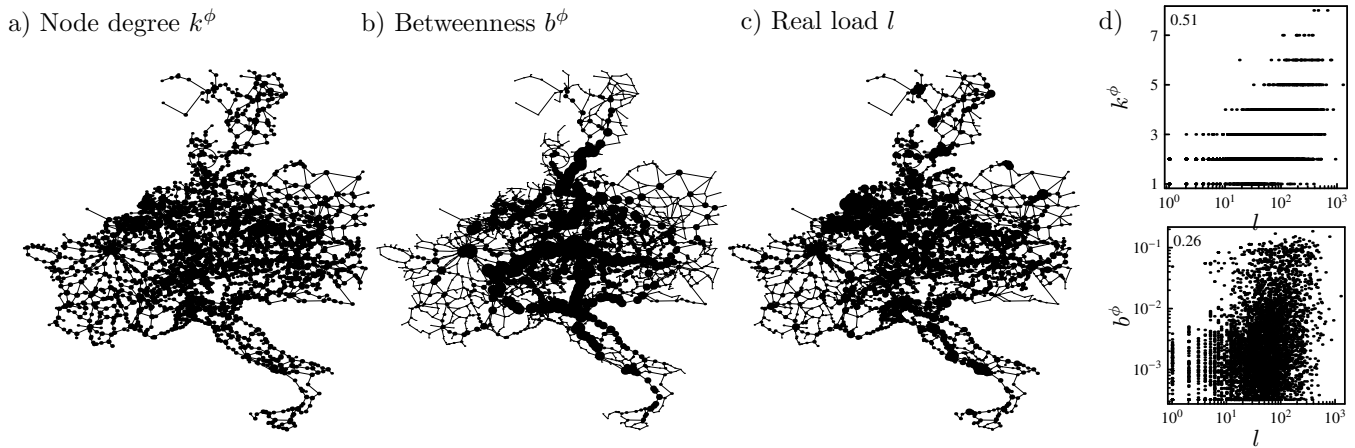}
 \\[-0.25cm]
 \caption{EU dataset (WA and CH yield similar results).
The first three figures present the physical layout of the node degree $k^\phi$~(a), betweenness $b^\phi$~(b) and real
load $l$~(c). The size of a node is proportional to the measured value. In (d) we present the scatter-plots of the node
degree $k^\phi$ (top) and betweenness $b^\phi$ (bottom) versus the real load~$l$.
In the top left corner of every plot we give the value of the corresponding Pearson's correlation coefficient.\\[-0.8cm]}
\label{fig:LoadComparison}
\end{figure*}
We study three examples of transportation networks, with sizes ranging from city to continent. As an example of a city,
we take the mass transportation system (buses, trams and metros) of Warsaw (WA), Poland. At a country level, we study
the railway network of Switzerland (CH). Finally we investigate the railway network formed by major trains and stations
in most countries of central Europe (EU). The basic characteristics of these networks can be found in
Table~\ref{tab:Data} and in Fig.~\ref{fig:Flows}. All physical topologies are connected, planar or close to planar,
with the diameter $d^\phi$ in the order of $O(\sqrt{N})$ (the diameter of a graph is the length of the longest of all
possible shortest paths), and node degree distributions decaying exponentially (the degree of a node is the number of
edges incident on this node). These features are common to many physical transportation graphs, such as a road network
or a railway system. The logical graphs are strikingly different. They are sparse and have multiple components, among
which many isolated nodes. The degree distributions of the logical graphs are \emph{right-skewed}, meaning that there
is a small number of nodes with very high degree. This is confirmed by the almost linear shape of the distribution in
the log-log scale plot shown in Fig.~\ref{fig:Flows}b; a fully linear shape would indicate a power-law (a heavy-tailed
distribution).
Similar right-skewed distributions are observed for the weights of edges in the logical graphs (Fig.~\ref{fig:Flows}c).
In Fig.~\ref{fig:Flows}d, we compare the length distribution of real traffic flows with the length distribution of
all-to-all shortest paths. The former is very much left-skewed, which means that the real flows tend to be local.

Knowing the topologies and the mapping of both layers, we can easily compute the load of a node with
formula~(\ref{eq:node_load}). For comparison purposes, we present below two \emph{load estimators} based exclusively on
the physical graph~$G^\phi$. For load estimators we take two metrics known in social networks as centrality measures;
they are used to assess the importance of nodes. Our first metric is \emph{node degree}~$k^\phi$. It seems natural that
the nodes with a high degree carry more traffic than the less connected nodes. Our second metric is
\emph{betweenness}~$b^\phi$~\cite{Freeman77}. The betweenness of a vertex $v$ is the fraction of shortest paths between
all pairs of vertices in a network, that pass through $v$. If there are more than one shortest path between a given
pair of vertices, then all such paths are taken into account with equal weights summing to one. Betweenness aims at
capturing the amount of information passing through a vertex. Indeed, many authors take betweenness as a measure of
load either directly~\cite{Goh01,Holme02,Motter02,Szabo02,Bollobas04,Zhao04}, or with slight
modifications~\cite{AlbertUSAPowerGrid,Moreno04,Tadic04}

In Fig.~\ref{fig:LoadComparison} we compare the distribution of the real load with its two estimators: node degree and
betweenness. The geographical patterns formed by the three metrics differ substantially (see
Fig.~\ref{fig:LoadComparison}abc). To quantify these differences, in Fig.~\ref{fig:LoadComparison}d we present the
scatter plots of these two estimators versus the real load $l$. The correlations between them are very low, which is
confirmed by low values of the corresponding Pearson's coefficients (top left corner of every plot). For instance, for
the value of load $l\simeq 10^2$, the corresponding values of betweenness $b^\phi$ cover more than two orders of
magnitude. Surprisingly, contrary to the commonly admitted view, the node degree approximates the real load better than
betweenness.

Why do load estimators fail to mimic the real load pattern? Are there some fundamental reasons behind this? The layered
view of the system is very helpful in answering these questions. First, observe that the ways we compute node degree,
betweenness and real load, can be unified by recasting the first two in the two-layered setting. Indeed, both the node
degree and the betweenness can be computed as the node load~(\ref{eq:node_load}) in two-layered systems with specific
logical topologies mapped on the physical graph~$G^\phi$ using shortest paths. We denote these specific logical graphs
by $G^\lambda_{\!k\!^\phi}$ and  $G^\lambda_{\!b^\phi}$, for the node degree~$k^\phi$ and the
betweenness~$b^\phi$, respectively. They are defined as follows.\\
In the case of the node degree, pick~$G^\lambda_{\!k\!^\phi}=G^\phi$: the logical graph is identical to the physical
graph~$G^\phi$. Hence the mapping of $G^\lambda_{\!k\!^\phi}$ on $G^\phi$ reduces trivially to single hop traffic
flows, and~(\ref{eq:node_load}) boils down to $l(v^\phi)=k^\phi(v^\phi)$.\\
For the betweenness, $G^\lambda_{\!b^\phi}$ is an unweighted and complete (fully connected) graph. Indeed, the
definition of betweenness requires to find shortest paths between every possible pair of vertices. Note that the
mapping defined by betweenness splits the path (and its weight) if there are more than one shortest path, whereas the
shortest-path mapping simply returns one of them. However, in large graphs this difference is negligible, especially if
the shortest-path algorithm picks one of the possible paths at random.\\
The same two-layered methodology can therefore be used to compute node degree, betweenness and real load. Moreover, in
all three cases we use the same physical graph $G^\phi$ and a mapping that follows the shortest path~\footnote{
The real-life flows almost always coincide with shortest paths connecting their end-nodes.}. Consequently, all the
differences between the three metrics are completely captured by the logical graphs $G^\lambda_{\!k\!^\phi}$,
$G^\lambda_{\!b^\phi}$ and $G^\lambda$. We compare them in Table~\ref{tab:LogGraphs}. The
graph~$G^\lambda_{\!k\!^\phi}$ is moderately dense, planar, unweighted, with the degree distribution decaying
exponentially. The edge length, counted in the number of hops in the mapping of this edge, is  equal to one for all
edges of $G^\lambda_{\!k\!^\phi}$. In contrast, the graph $G^\lambda_{\!b^\phi}$ is an unweighted and complete graph,
which means it is very dense, with every node of degree equal to $N\!-\!1$. In $G^\lambda_{\!b^\phi}$ we find both
short and long edges; their distribution is bell-shaped, as shown by the ``all-to-all'' curve in Fig.~\ref{fig:Flows}d.
Finally, the real-life logical graph $G^\lambda$ is sparse, weighted and has rather local edges (see the ``real'' curve
in Fig.~\ref{fig:Flows}d). Moreover, the node degree and edge weight distributions of $G^\lambda$ are both very much
right-skewed.

\begin{table}[!t]
\begin{small}
\begin{tabular}{|c|c|c|c|}
 \hline
 Property &
 \raisebox{0.5cm}{\normalsize $G^\lambda_{\!k\!^\phi}$}\hspace{-0.6cm}
 \begin{minipage}[c]{1.9cm} \includegraphics[width=0.85\textwidth]{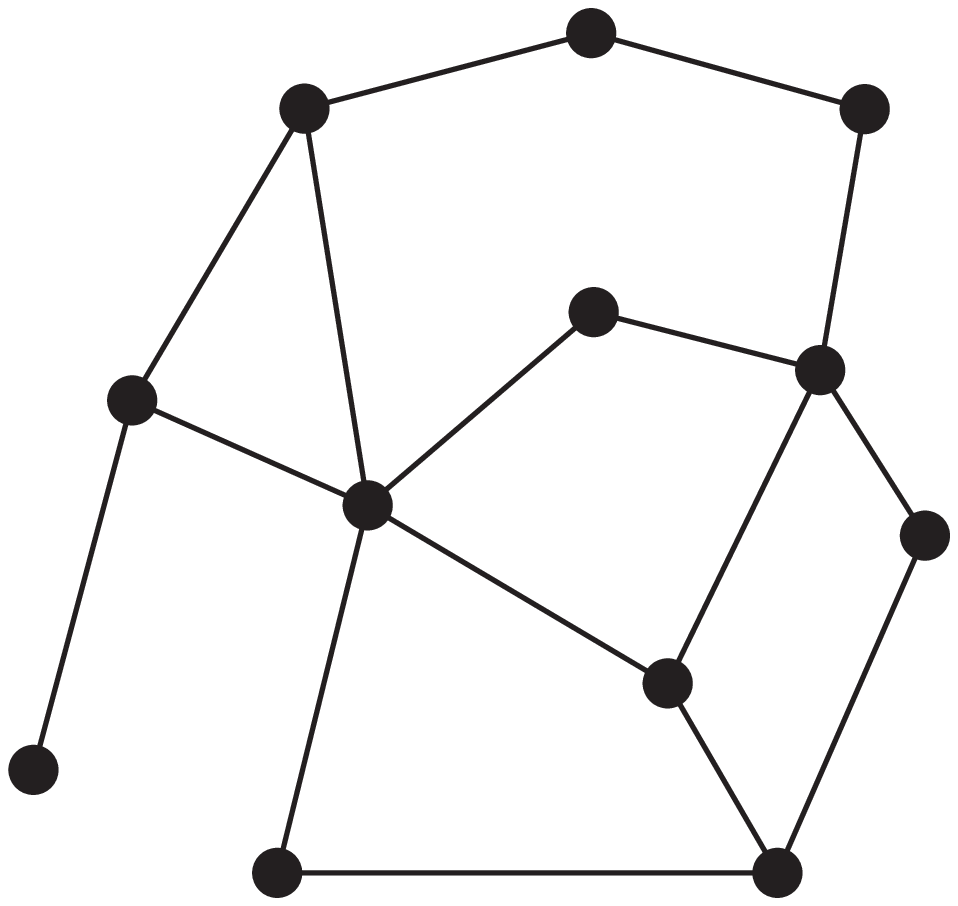}\end{minipage}&
 \raisebox{0.5cm}{\normalsize $G^\lambda_{\!b^\phi}$}\hspace{-0.6cm}
 \begin{minipage}[c]{1.9cm} \includegraphics[width=0.9\textwidth]{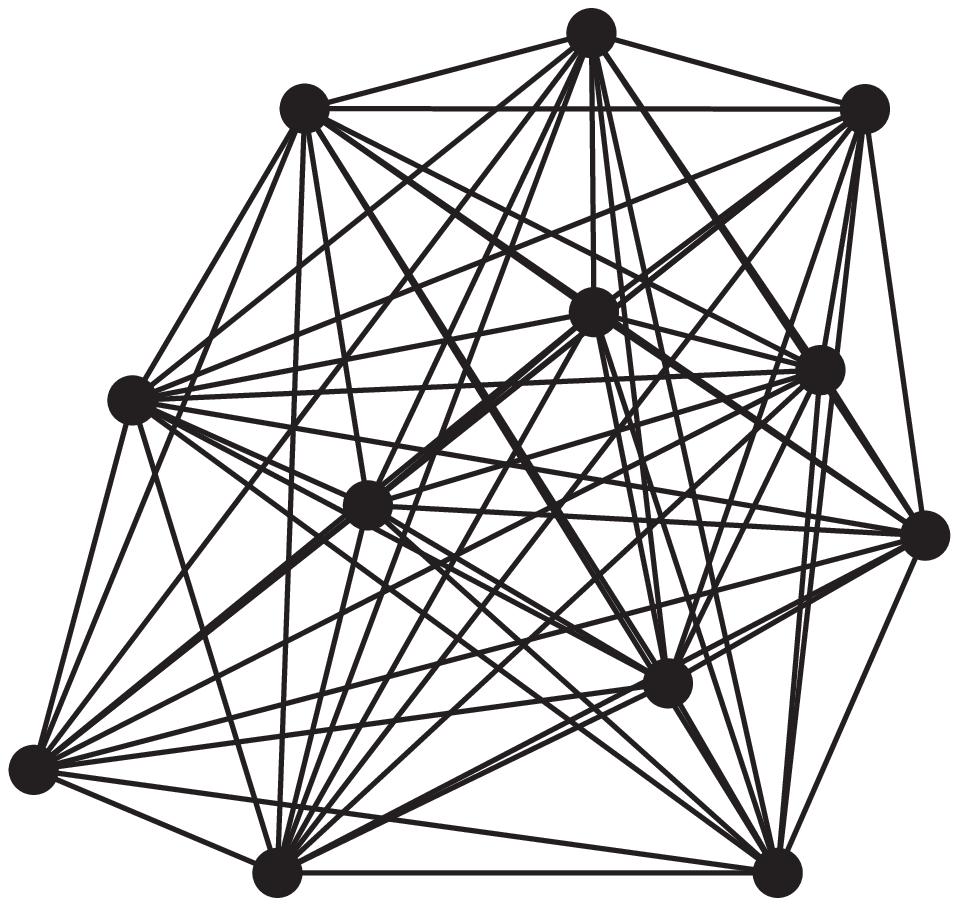}\end{minipage}&
  \raisebox{0.5cm}{\normalsize $G^\lambda$}\hspace{-0.6cm}
 \begin{minipage}[c]{1.9cm} \includegraphics[width=0.9\textwidth]{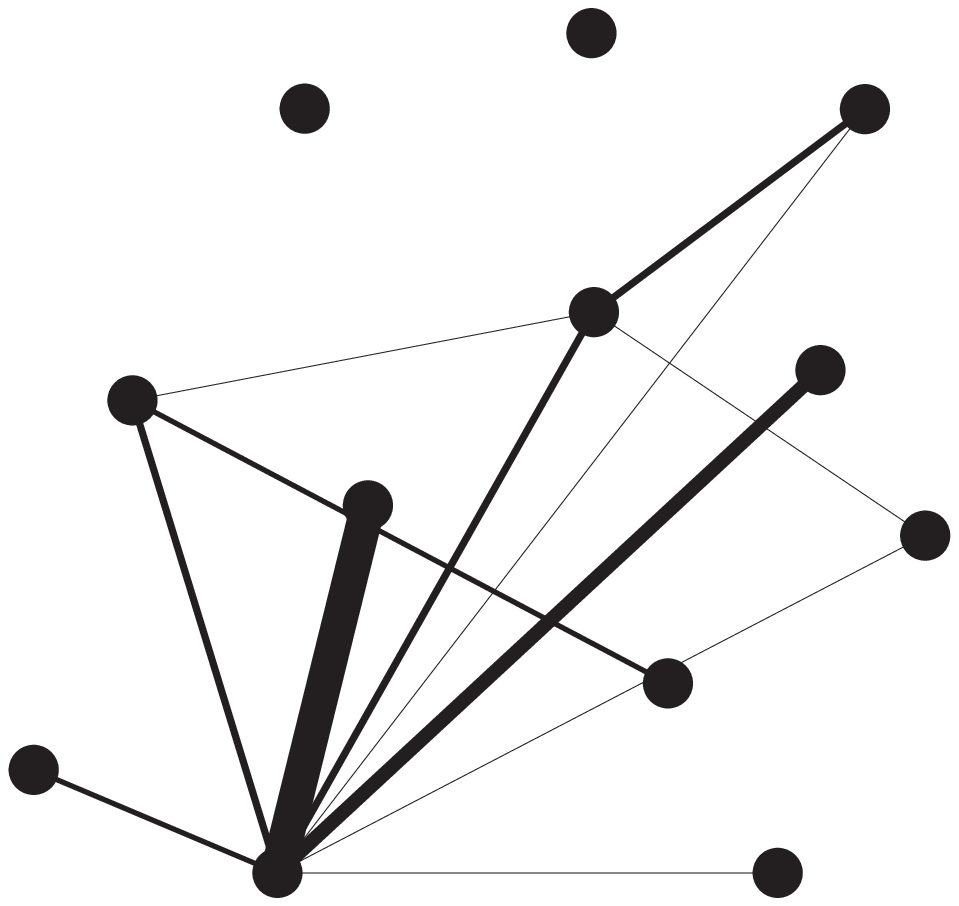}\end{minipage}\\
  \hline
  $\mid E^\lambda\mid$ & $=\mid E^\phi\mid$ & $=N(N-1)/2$ & $<\mid E^\phi\mid$\\
  Planar & Yes & No &No\\
  Weights $w(e^\lambda)$& = 1& = 1 & Right-skewed\\
  Degrees $k(v^\lambda)$& Exponential & $= N-1$ & Right-skewed\\
  Edge lengths& = 1 & Bell-shaped & Exponential\\
   \hline
\end{tabular}
\caption{The properties of the logical graphs $G^\lambda_{\!k\!^\phi}$, $G^\lambda_{\!b^\phi}$ and $G^\lambda$.
``Edge length'' is the number of hops in the mapping of the edge on the physical graph.\\[-0.7cm]} \label{tab:LogGraphs}
\end{small}
\end{table}

\pagebreak[4] There are thus a number of fundamental differences between the three logical
graphs~$G^\lambda_{\!k\!^\phi}$, $G^\lambda_{\!b^\phi}$ and $G^\lambda$. They explain why the node degree and
betweenness fail to mimic the real load distribution. We expect to observe similar differences in other fields. For
instance, the logical graph representing the traffic in the Internet shares many properties with the logical graphs of
transportation systems studied here. In particular, in the Internet, the distribution of intensity of traffic flows
(which is, in this paper, equivalent to the edge weights in the logical graph) was shown to be
heavy-tailed~\cite{Elephants,Meiss05}.
This is known in the field as ``the elephants and mice phenomenon''~\cite{Elephants}, where a small fraction of flows
is responsible for carrying most of the traffic. Moreover, the number of flows originating from a given node (which is,
in this paper, equivalent to the node degree in the logical graph), was also shown to follow a power-law
distribution~\cite{Meiss05}.

To summarize, we have introduced a framework for studying complex systems in which we distinguish graphs on two or more
layers. We have shown on the example of transportation networks how the layered view can facilitate the description,
comparison and analysis of such systems. Our work represents only a fraction of the possibilities in this area. For
example, the layered perspective can completely change our view of the error and attack tolerance of considered
systems. It would be also interesting to study how the properties of the topologies at different layers affect the
interactions between the layers.

\end{document}